\documentclass[aps,prl,reprint,showpacs,groupedaddress]{revtex4-1}

\usepackage{graphicx}
\usepackage{dcolumn}
\usepackage{amssymb}
\usepackage{bm}

\begin{document}


\title{Thermal Pure Quantum States at Finite Temperature}


\author{Sho Sugiura}
\email[]{sugiura@ASone.c.u-tokyo.ac.jp}
\author{Akira Shimizu}
\email{shmz@ASone.c.u-tokyo.ac.jp}
\affiliation{Department of Basic Science, 
University of Tokyo, 3-8-1 Komaba, Meguro, 
Tokyo 153-8902, Japan}


\date{\today}

\begin{abstract}
An equilibrium state can be represented by a
pure quantum state, which we call a thermal pure quantum (TPQ) state.
We propose a new TPQ state and a simple method of obtaining it.
A single realization of the TPQ state 
suffices for calculating all
statistical-mechanical properties, including correlation functions and
genuine thermodynamic variables, of a quantum system at finite temperature.
\end{abstract}


\maketitle

The possibility of extracting statistical-mechanical 
information from a pure quantum state has been intensively discussed in the context of the foundation of statistical mechanics 
 \cite{Popescu,Lebowitz,SugitaJ,Reimann}.
As we shall demonstrate here, it also has a potential significance for 
a new formulation of 
statistical mechanics, 
and for a novel calculation technique.

As an illustration, let us 
consider a closed quantum system composed of $N$ spins, 
which is enclosed by adiabatic walls.
In the ensemble formulation,
its equilibrium properties are described by the microcanonical 
ensemble, which is specified by $E$ (energy), $N$, and so on. 
The corresponding subspace (energy shell) 
in the Hilbert space $\mathcal{H}_N$ is denoted
by $\mathcal{E}_{E,N}$.
Let us consider a random vector 
$
| \psi \rangle = 
{\sum_\nu}^{\!\!\! \prime} \ c_\nu |\nu \rangle
$
in $\mathcal{E}_{E,N}$, 
where
$\{ |\nu \rangle \}_\nu$ is an arbitrary orthonormal basis set
of $\mathcal{E}_{E,N}$, 
${\sum_\nu}^{\!\!\! \prime}~$ denotes the sum over this basis,
and $\{ c_\nu \}_\nu$ is 
a set of random complex numbers drawn uniformly
from the unit sphere ${\sum_\nu}^{\!\!\! \prime} \ |c_\nu|^2 =1$ in the 
complex space of dimension $\dim \mathcal{E}_{E,N}$.
It was shown in Refs.~\cite{Popescu,Lebowitz,SugitaJ,Reimann} 
that almost every such vector 
gives the correct equilibrium values of 
a certain class of observables $\hat{A}$
by $\langle \psi | \hat{A} | \psi \rangle$.
This property was proved in Refs.~\cite{Popescu,Lebowitz}
for observables of a subsystem, which is much smaller than the whole system.
The case of general observables, including
observables of the whole system (such as the total magnetic moment and its fluctuation), 
was analyzed in Refs.~\cite{SugitaJ,Reimann}.
It was shown that the above property holds {\em not} for all observables
but for observables that are low-degree polynomials 
(i.e., their degree $\ll N$) of local operators \cite{SugitaJ}.
We here call such observables {\em mechanical variables}.
We 
assume that all mechanical variables are 
normalized in such a way that 
they are dimensionless.

For conceptual clarity, we call {\em generally} a pure quantum state that 
represents an equilibrium state a 
{\em thermal pure quantum state} (TPQ state).
Stating more precisely for the case where 
a state $|\psi \rangle$ has random variables 
(such as the random vector 
discussed above), 
we call $|\psi \rangle$ a TPQ state 
if for an arbitrary positive number $\epsilon$
\begin{equation}
{\rm P}( 
| 
\langle \psi | \hat{A}| \psi \rangle
-
\langle \hat{A} \rangle^{\rm eq}_{E, N}
|
\geq \epsilon
) 
\leq
\eta_\epsilon (N)
\label{eq:Pto0}
\end{equation}
for every mechanical variable $\hat{A}$.
Here, 
${\rm P}(x)$ denotes the probability of event $x$,
$\langle \cdot \rangle^{\rm eq}_{E, N}$ denotes the ensemble average,
and 
$\eta_\epsilon (N)$
is a function (of $N$ and $\epsilon$) which vanishes as $N \to \infty$.
%
%
The above inequality means that for large $N$ 
getting a {\em single} realization of 
a TPQ state is sufficient, with high probability, 
for evaluating 
equilibrium values 
of mechanical variables. 
The vector
$
{\sum_\nu}^{\!\!\! \prime} \ c_\nu |\nu \rangle
$
of Refs.~\cite{Popescu,Lebowitz,SugitaJ,Reimann} 
is a TPQ state.
However, important problems remain to be solved.
Most crucially, 
{\em genuine thermodynamic variables}, 
such as the entropy and temperature,
cannot be calculated as $ \langle \psi | \hat{A} | \psi \rangle$
because they are not mechanical variables \cite{vNS}.
Moreover, 
one needs to prepare a basis $\{ |\nu \rangle \}_\nu$ of $\mathcal{E}_{E,N}$ 
to construct ${\sum_\nu}^{\!\!\! \prime} \ c_\nu |\nu \rangle$.
Since this is a hard task, such a TPQ state 
is hard to obtain.

In this Letter, we resolve these problems
by proposing a new TPQ state, 
a novel method of constructing it,
and new formulas for obtaining genuine thermodynamic variables.
This novel formulation of statistical mechanics
enables one to calculate
{\em all} variables of statistical-mechanical interest
at {\em finite temperature},
from only a {\em single} realization of the TPQ state.
We also show that 
this formulation is very useful for practical calculations.
%
%

{\em New TPQ state -- }
We consider a discrete quantum system composed of $N$ sites, 
which is described by a Hilbert space $\mathcal{H}_N$ 
of dimension $D=\lambda^N$,
where $\lambda$ is a constant of $O(1)$.
[For a spin-1/2 system, $\lambda=2$.]
Our primary purpose is to obtain results in the thermodynamic limit: 
$N \rightarrow \infty$ while $E/N$ is fixed. 
Therefore, we hereafter use quantities per site,
$\hat{h} \equiv \hat{H}/N$ (where $\hat{H}$ denotes the Hamiltonian), 
$u \equiv E/N$, and $(u; N)$ instead of $(E,N)$.
[We do not write explicitly 
variables other than $u$ and $N$, such as a magnetic field.]
We assume that the system is consistent with thermodynamics
in the sense that 
the density of states $g(u;N)$ behaves as \cite{en:Boltzmann}
\begin{equation}
g(u;N)=  
\exp[Ns(u;N)], 
\
\beta'(u;N)
\leq 0.
\label{eq:TDconsistency}
\end{equation}
Here, $s(u;N)$ is the entropy density,
which converges to the $N$-independent 
one $s(u;\infty)$ as $N\rightarrow \infty$,
$\beta(u;N) \equiv \partial s(u;N)/\partial u$
is the inverse temperature,
and
$\beta'
\equiv
\partial \beta /\partial u
$.
These conditions are satisfied, for example,
by spin models and the Hubbard model.
Since $D$ is finite, 
$\beta$ may be positive and negative in  
lower- and higher-energy regions,
respectively.
We here consider the former region.

%
%
%

We propose the following TPQ state and 
the procedure for constructing it. 
First, take a random vector 
$
| \psi_{0} \rangle 
\equiv
\sum_i c_i | i \rangle
$
from the {\em whole} Hilbert space $\mathcal{H}_N$.
Here, $\{|i\rangle\}_i$ is an arbitrary orthonormal basis of $\mathcal{H}_N$, 
and $\{c_i\}_i$ is a set of random complex numbers drawn uniformly
from the unit sphere $\sum|c_i|^2=1$ of the $D$-dimensional complex space. 
Note that this construction of random vectors is independent of the choice of the orthonormal basis $\{|i\rangle\}_i$.
One can therefore use a trivial basis such as a set of product states.
Hence, $| \psi_{0} \rangle$ can be generated easily.
On ther whole, 
the amplitude is almost equally distributed over {\em all} the energy eigenstates 
in this state
(as is easily seen by choosing the eigenstates of $\hat{h}$ as the basis $\{|i\rangle\}_i$).
Thus, the distribution of energy in $|\psi_0\rangle$ is proportional to $g(u;N)$.
We wish to modify this distribution into another distribution $r_k(u;N)$ 
which has a peak at an desired energy.
This is easily done by operating a suitable polynomial of $\hat{h}$
onto $|\psi_0 \rangle$ as we shall see below. 
[Operating  $\hat{h}$ onto a vector is much easier 
than diagonalizing $\hat{h}$.]
We denote the minimum and the maximum eigenvalues of $\hat{h}$
by $e_{\rm{min}}$ and $e_{\rm{max}}$, respectively.
Take a constant $l$ of $O(1)$ 
such that $l \ge e_{\rm max}$. 
Starting from $| \psi_{0} \rangle$, calculate
\begin{eqnarray}
u_{k} &\equiv& \langle \psi_{k} | \hat{h} | \psi_{k} \rangle,
\label{eq:uk}
\\
| \psi_{k+1} \rangle 
&\equiv&
( l - \hat{h} ) | \psi_{k} \rangle
/
\| ( l - \hat{h} ) | \psi_{k} \rangle \|
\end{eqnarray}
iteratively for $k=0, 1, 2, \cdots$. 
From Eq.~(\ref{Nb=k/L}) below, 
$u_{0}$ corresponds to $\beta=0$, 
i.e., $g(u;N)$ takes the maximum at $u=u_{0}$.
We will 
also show that $u_{k}$ decreases gradually down to $e_{\rm min}$ as $k$ is increased, 
i.e., 
$
u_{0} > u_{1} > \cdots \geq e_{\rm{min}} 
$.
One may terminate the iteration 
when $u_{k}$ gets low enough for one's purpose.
We denote $k$ at this point by $k_{\rm term}$. 
We will show that
$k_{\rm term} = O(N)$ at finite temperature,
and that 
the states
$ 
	| \psi_{0} \rangle, | \psi_{1} \rangle, 
	\cdots,
	| \psi_{k_{\rm term}} \rangle
$ 
become a series of TPQ states corresponding to various energy densities,
$u_{0}, u_{1}, \cdots, u_{k_{\rm term}}$.
%
%
Hence, the equilibrium value 
of an arbitrary mechanical variable $\hat{A}$ is obtained as
$\langle \psi_{k} | \hat{A} | \psi_{k} \rangle$, 
as a function of $u_{k}$. 
For each realization of $\{ c_i \}_i$, 
a series of realizations of TPQ states is obtained.
We will show that the dependence of 
$\langle \psi_{k} | \hat{A} | \psi_{k} \rangle$ on $\{ c_i \}_i$ is
exponentially small in size $N$ as $N$ increases.
Therefore, 
{\em only a single realization
suffices for getting
a fairly accurate value. }
When better accuracy is required, 
one can take the average over many realizations. 

We now show that the states obtained with the above procedure are TPQ states.
Since $| \psi_{0} \rangle$ is independent of the choice of the basis, 
we take 
the set of energy eigenstates $\{ | n \rangle \}_n$
as 
$\{ | i \rangle \}_i$ 
in order to see properties of $| \psi_k \rangle$
(although we never use such a basis in practical calculations).
After $k$-times multiplication of $l-\hat{h}$, 
$ | \psi_{0} \rangle
=
\sum_n c_n | n \rangle
$
turns into 
\begin{equation}
| \psi_{k} \rangle  
\propto 
(l-\hat{h})^k | \psi_{0} \rangle
=  
\sum_{n} c_{n} (l-e_{n})^k |n \rangle,
\label{psi=Sdn}
\end{equation}
where $\hat{h} |n \rangle = e_n |n \rangle$. 
Let us examine how the energy density $u$ distributes in this state.
The (unnormalized) distribution function of $u$ is given by
$
r_k(u;N)
\equiv
\delta_r^{-1} {\sum_n}^{\!\! \prime \prime} |c_{n}|^2 (l-e_{n})^{2k}
$,
where $\delta_r = o(1)$ and
the sum is taken over $n$ such that $e_n$ lies in 
a small interval $[u-\delta_r /2, u+\delta_r /2)$.
Since the density of states $g(u;N)$ is exponentially large in size $N$, 
$r_k(u;N)$ converges (in probability) exponentially fast
to its average. Hence, 
\begin{equation}
r_k(u;N)
= D^{-1}  \exp[N \xi_{\kappa}(u;N)],
\label{Sd=e^xi}
\end{equation}
where $\xi_{\kappa}(u;N) \equiv s(u;N) + 2\kappa \ln (l-u)$ with $\kappa \equiv k/N$.
Hereafter we often denote $k$ dependence by $\kappa$, 
e.g., we express $u_k$ as $u_{\kappa}$.
Note that $\xi_{\kappa}(u;N)$ does not depend on $\{ c_i \}_i$,
because the dependence vanishes
when we have dropped negligible terms in Eq.~(\ref{Sd=e^xi}). 
$\xi_{\kappa}(u;N)$ takes the maximum at $u^{\ast}_{\kappa}$ which satisfies
\begin{eqnarray}
\beta(u^{\ast}_{\kappa};N) = 2 \kappa / (l-u^{\ast}_{\kappa}).
\label{Nb=k/L}
\end{eqnarray}
Since $\beta(u^*_\kappa; N)$ and $l-u^{\ast}_{\kappa}$ are $O(1)$, 
we find $\kappa = O(1)$, and hence $k = O(N)$.
Expanding $\xi_{\kappa}(u;N)$ around $u^{\ast}_{\kappa}$, 
and noticing 
\begin{eqnarray}
\xi''_{\kappa} 
\equiv \partial^2 \xi_{\kappa}/\partial u^2 
= \beta'(u^{\ast}_{\kappa};N) - 2\kappa/(l-u^{\ast}_{\kappa})^2 < 0
\nonumber
\end{eqnarray}
from Eq.~(\ref{eq:TDconsistency}), 
we get 
$
\xi_{\kappa}(u;N) 
=
\xi_{\kappa}(u^{\ast}_{\kappa};N) 
- |\xi''_{\kappa}| (u-u^{\ast}_{\kappa})^2 / 2
+ \xi'''_{\kappa} (u-u^{\ast}_{\kappa})^3 / 6
+ \cdots
$.
Here, 
$\xi'''_{\kappa} \equiv \partial^3 \xi_{\kappa}/\partial u^3
=\beta''(u^{\ast}_{\kappa};N) - 4\kappa/(l-u^{\ast}_{\kappa})^3$.
Hence, $r_k(u;N)$ behaves almost as the Gaussian distribution, peaking at 
$u=u^*_\kappa$, 
with the vanishingly small variance $1/N |\xi''_{\kappa}|$.
Let us introduce the density operator 
$
\hat{\rho}_k \equiv 
(l-\hat{h})^{2k}
/{\rm Tr}(l-\hat{h})^{2k}
$, 
which has the same energy distribution $r_k(u;N)$.
In the ensemble formulation, 
$\hat{\rho}_k$ represents the equilibrium state specified by 
$(u_\kappa; N)$
because $r_k(u;N)$ has a sharp peak.
We call the ensemble corresponding to $\hat{\rho}_k$
the {\em smooth microcanonical ensemble} (because
the energy distribution is smooth).
In a way similar to those of Refs.~\cite{SugitaJ,Reimann}, 
we can show that for an arbitrary positive number $\epsilon$
\begin{eqnarray}
{\rm P}\Big( 
\Big| \langle \psi_{k} | \hat{A} | \psi_{k} \rangle
&-&
{\rm Tr}[\hat{\rho}_k \hat{A}] \Big|
\geq \epsilon
\Big) 
\leq 
{ \| \hat{A} \|^2 r_k(e_{\rm min};N) \over \epsilon^2  r_k(u_\kappa^*;N)},
\quad
\label{DeltaA}
\\
\overline{\langle \psi_{k} | \hat{A} | \psi_{k} \rangle}
&=&
{\rm Tr}[\hat{\rho}_k \hat{A}]
\label{<A>}
\end{eqnarray}
for every mechanical variable $\hat{A}$.
Here, 
$\| \cdot \|$ denotes the operator norm \cite{norm}, and
the overline represents the random average.
With increasing $N$, 
$\|\hat{A} \|^2$ grows at most as a low-degree polynomial of $N$,
whereas 
$r_k(e_{\rm min};N)/r_k(u_\kappa^*;N)$ 
decreases exponentially at finite temperature 
(i.e., for $u_\kappa^* > e_{\rm min}$).
Therefore, {\em $| \psi_{k} \rangle$ is a TPQ state
for the smooth microcanonical ensemble}. 

{\em Genuine thermodynamic variables -- }
One might think it impossible to obtain genuine thermodynamic 
variables 
like the temperature and entropy by only manipulating pure quantum states.
However, 
our new TPQ state makes it possible.
In fact, 
by substituting $u_{\kappa}$ for $u^{\ast}_{\kappa}$ in Eq.~(\ref{Nb=k/L}),
and using Eq.~(\ref{eq:u'}) below, 
we obtain 
\begin{eqnarray}
\beta(u_{\kappa};N) = 2 \kappa / (l-u_{\kappa}) + O(1/N).
\label{Nb=k/L+O(1/N)}
\end{eqnarray}
This gives $\beta(u_{\kappa};N)$, with an error of $O(1/N)$, 
as a function of $u_{\kappa}$
[because 
$\kappa$ and $l$ are known parameters]. 
That is, one obtains the temperature of the equilibrium state
specified by $(u_{\kappa};N)$ just by calculating 
 $u_{\kappa}$ with Eq.~(\ref{eq:uk}).

We can also obtain formulas with less errors.
For example, 
%
using Eq.~(\ref{Sd=e^xi}) and the expansion of $\xi_\kappa(u;N)$, we have
\begin{equation}
u^{\ast}_{\kappa} = u^{\bullet}_{\kappa} + O(1/N^2),
\
u^{\bullet}_{\kappa} \equiv 
u_{\kappa} - \xi'''_{\kappa}/2N{\xi''_{\kappa}}^2.
\label{eq:u'}
\end{equation}
Substituting 
$u^{\bullet}_{\kappa}$ for $u^{\ast}_{\kappa}$ in Eq.~(\ref{Nb=k/L}),
we get a better formula 
\begin{equation}
\beta(u^{\bullet}_{\kappa}; N)
=
2 \kappa / (l - u^{\bullet}_{\kappa})
+ O(1/N^2).
\label{eq:betaN_better}\end{equation}
One can evaluate 
$\xi''_{\kappa}$ and $\xi'''_{\kappa}$ 
easily by calculating 
$
\langle \psi_{k} | 
(\hat{h}  - u_{\kappa} )^2
| \psi_{k} \rangle
= 1/N|\xi''_{\kappa}| + O(1/N^2)
$
and
$
\langle \psi_{k} | 
(\hat{h}  - u_{\kappa} )^3
| \psi_{k} \rangle
=
\xi'''_{\kappa}/N^2 |\xi''_{\kappa}|^3 + O(1/N^3)
$.
Hence, 
using formula (\ref{eq:betaN_better}), 
one obtains  
$\beta(u; N)$ (for $u = u^{\bullet}_0, u^{\bullet}_1,\cdots$) 
with an error of $O(1/N^2)$.
In a similar manner, we can obtain formulas 
whose errors are of even higher order of $1/N$.


However, $\beta(u; N)$ is the inverse temperature of a {\em finite} system,
whereas we are most interested in its thermodynamic limit $\beta(u; \infty)$.
In general, the difference $|\beta(u; N) - \beta(u; \infty)|$ decays 
not so quickly as $O(1/N^2)$.
To obtain an even better formula for $\beta(u; \infty)$,
we consider $C$ identical copies of the $N$-site system.
We denote quantities of this $CN$-site system by tilde, 
such as 
$|\tilde{\psi}_{0} \rangle 
\equiv |\psi_{0} \rangle^{\otimes C}$.
The state $| \tilde{\psi}_{\tilde{k}} \rangle$ 
is given by
$ 
| \tilde{\psi}_{\tilde{k}} \rangle 
\propto (\tilde{l}-\tilde{h})^{C \tilde{k}} | \tilde{\psi}_{0} \rangle,
$ 
where 
$
\tilde{h} 
\equiv 
(
\hat{H} \otimes \hat{1}^{\otimes (C-1)} 
+ \hat{1} \otimes \hat{H} \otimes \hat{1}^{\otimes (C-2)} 
+ \cdots + 
\hat{1}^{\otimes (C-1)} \otimes \hat{H}
)/CN$. 
In the limit of $C \rightarrow \infty$, 
$\tilde{u}_{\tilde{\kappa}}$ approaches 
the canonical average of $u$ in a single copy 
with inverse temperature $\tilde{\beta}(\tilde{u}_{\tilde{\kappa}};\infty)$.
At the point where 
$\tilde{\beta}(\tilde{u}_{\tilde{\kappa}};\infty)
=\beta(u^{*}_{\kappa}; N)$
is satisfied,
we can estimate this canonical average,
which is denoted by $\tilde{u}^{\rm c}_\kappa$, 
in the same manner as Eq.~(11).
Then, we get 
$\tilde{u}^{\rm c}_\kappa = \tilde{u}^{\bullet}_\kappa +O(1/N^2)$,
where
\begin{equation}
\tilde{u}^{\bullet}_\kappa
\equiv
u^{\bullet}_{\kappa}
+\frac{\xi'''_{\kappa}+ 4\kappa/(l-u^{\bullet}_\kappa)^3}
{2N [\xi''_{\kappa} + 2\kappa/(l-u^{\bullet}_{\kappa})^2]^2}.
\label{u=u-z/g+b''/b'}
\end{equation}
We thus find
\begin{equation}
\tilde{\beta}(\tilde{u}^{\bullet}_\kappa;\infty) 
= 
2\kappa/(l-u^{\bullet}_\kappa)
+O(1/N^2),
\label{eq:beta_infty}\end{equation}
which gives the inverse temperature 
$\tilde{\beta}(u;\infty)$ (for $u=\tilde{u}^{\bullet}_0,\tilde{u}^{\bullet}_1,\cdots$)
of an infinite system
composed of an infinite number of $N$-site systems.
We expect that $\tilde{\beta}(u;\infty)$ is much closer to $\beta(u;\infty)$
than $\beta(u;N)$,
because
information of $\xi(u;N)$ in the whole spectrum range of $u$
is included in 
$\tilde{\beta}(u;\infty)$.
[By contrast, only the information at the peak of 
$\xi(u;N)$ is included in $\beta(u;N)$.]
This will be confirmed later by numerical computation.

We can also obtain the entropy density $s$ 
as a function of $u$ and $h_z$,  
by integrating $\beta$ over $u$ 
and $\beta m_z$ over $h_z$. 
For example, for an arbitrarily fixed value of $h_z$, we have
\begin{equation}
s(u_{2p}) - s(u_{2q})
=
\sum_{\ell = p}^{q-1} 
v(u^{\bullet}_{2\ell},u^{\bullet}_{2\ell+1},
u^{\bullet}_{2\ell+2})
+O({1 \over N^2}).
\label{s=Sum beta}
\end{equation}
by generalizing Simpson's rule.
Here, $u$ stands for 
$(u;N)$ or $(u;\infty)$,
$p$ and $q$ are integers, 
and
$v(x,y,z)
\equiv
(x-z) \{ \beta(x)+\beta(z) \}/2
-
(x-z)^2
[x \{ \beta(z)-\beta(y) \}+y \{ \beta(x)-\beta(z) \}
+z \{ \beta(y)-\beta(x) \}]
/6(x-y)(y-z)
$.
We have also developed another method of obtaining $s$, 
in which $g(u;N)$ 
is directly evaluated from the 
inner products
among different realizations of a TPQ state \cite{gakkai}.

To sum up,
one can obtain a series of TPQ states and values of all variables of statistical-mechanical 
interest, by preparing a random vector 
and simply applying $(l - \hat{h})$ iteratively.	
That is, 
we have established a new formulation of statistical mechanics,
whose fundamental formulas are 
Eqs.~(\ref{psi=Sdn}) and (\ref{Nb=k/L+O(1/N)}).

\begin{figure}
\begin{center}
\includegraphics[width=0.9\linewidth]{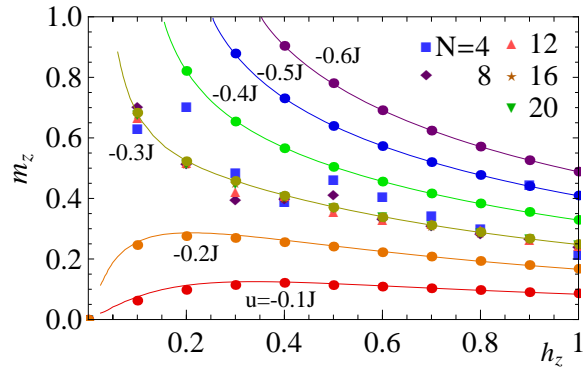}
\end{center}
\vspace{-6mm}
\caption{
Magnetization plotted against a magnetic field for $J=-1$.
Solid lines represent exact results for $N \to \infty$, 
for various values of the energy density $u$ \cite{Takahashi}. 
Circles denote results obtained with our formulation for $N=24$.
Results for $N=4$-$20$ are also shown for $u=-0.3J$.
}
\label{fig:hM}
\end{figure}

\begin{figure}
\begin{center}
\includegraphics[width=0.9\linewidth]{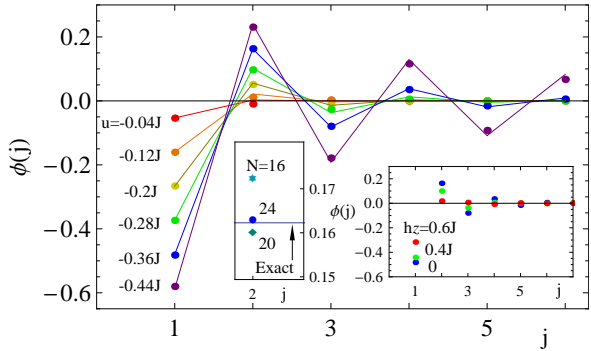}
\end{center}
\vspace{-6mm}
\caption{
Correlation function $\phi(j)$ plotted against $j$ for $J=+1$ and $h_z=0$.
Solid lines represent exact results for $N \to \infty$, 
for various values of $u$ \cite{Sato}. 
Circles denote results of our formulation for $N=24$.
(Left Inset) Results for $N=16$-$24$ at $j=2$ for $u=-0.36J$.
(Right Inset) $\phi(j)$ at finite $h_z$, 
obtained from a single realization of the TPQ state at $T \simeq 0.45J$.
}
\label{fig:Cor}
\end{figure}

{\em Numerical results -- }
Our formulation is easily implemented as a method of numerical 
computation.
We apply it to the one-dimensional Heisenberg model
in order to confirm the validity of the formulation.
We take
$
\hat{H} 
= \frac{J}{4} \sum_{i=1}^{N} 
[\hat{\bm{\sigma}}(i) \cdot \hat{\bm{\sigma}}(i+1) 
- h_z\hat{\sigma}_z(i)],
$ 
where $J=-1$ (ferromagnetic) or $+1$ (antiferromagnetic).
For $N \to \infty$, 
the exact results at finite temperature (i.e., $u > e_{\rm min}$)
have been 
derived for 
magnetization 
$
m_z
\equiv 
N^{-1} \sum_{i=1}^{N} 
\langle \sigma_z(i) \rangle^{\rm eq}_{u; N}
$ 
at all values of $u$ and $h_z$ \cite{Takahashi}, 
and for the correlation function 
$
\phi(j) \equiv 
N^{-1} \sum_{i=1}^{N} \langle \sigma_z(i) \sigma_z(i+j) \rangle^{\rm eq}_{u; N}
$ 
at all values of $u$ with $h_z=0$ \cite{Sato}.
They are plotted in Figs.~\ref{fig:hM} and \ref{fig:Cor}
by solid lines, where 
different colors correspond to different values of $u$.
We calculate the corresponding results using our formulation,
by performing numerical computation.
The results for $N=24$ are
plotted by circles, where each circle is obtained 
from a {\em single} realization of TPQ state.
According to Eq.~(\ref{DeltaA}), 
choice of the initial random numbers $\{ c_i \}_i$ 
has only an exponentially small effect on the results
at finite temperature. 
We have confirmed this fact 
by observing that 
the standard deviation, 
computed from ten realizations of a TPQ state for each data point,
is smaller than the radius of the circles of
these figures.

Results for other values of $N$ 
are plotted in Fig.~\ref{fig:hM} for $u=-0.3J$, 
and in the left insets of Fig.~\ref{fig:Cor} for $u=-0.36J$ at $j=2$.
It is seen that the $N$-dependence becomes fairly weak for $N \gtrsim 20$,
and that the results for $N=24$ agree well with the exact results.
As illustrated by this example, 
$N$ should be increased in our method until the 
variation of the results with increasing $N$ 
becomes less than the required accuracy.

We have also computed  
$\phi(j)$ at finite $h_z$ and $T$, 
for which {\em exact results are unknown}.
The results at $T \simeq 0.45J$ are plotted 
in the right inset of Fig.~\ref{fig:Cor}.

\begin{figure}
\begin{center}
\includegraphics[width=0.8\linewidth]{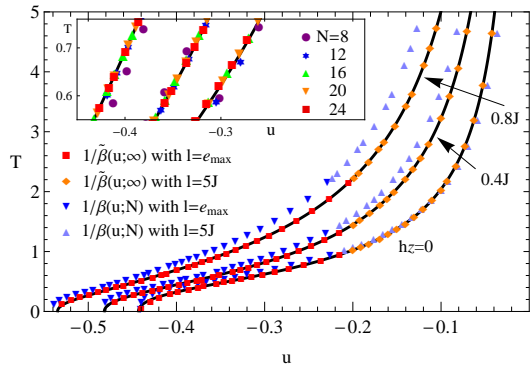}
\end{center}
\vspace{-6mm}
\caption{
Temperature $T$ plotted against $u$ for $J=+1$.
Solid lines represent exact results for $N \to \infty$, 
for various values of $h_z$ \cite{AF_ET}. 
Triangles and squares denote 
$1/\beta(u; N)$ (triangles) and 
$1/\tilde{\beta}(u; \infty)$ (squares)
for $N=24$, obtained with our  formulation. 
(Inset) 
$1/\tilde{\beta}(u; \infty)$ for $N=8$-$24$.
}
\label{fig:ET}
\end{figure}

For genuine thermodynamic variables, 
the exact result for $1/\beta(u; \infty)$ \cite{AF_ET} is 
plotted  by solid lines in Fig.~\ref{fig:ET}.
Corresponding results for 
$1/\beta(u; N)$ and $1/\tilde{\beta}(u; \infty)$, 
obtained with our method with $N=24$, are 
plotted by triangles and squares respectively, 
where each point is obtained 
from a {\em single} realization of the TPQ state.
[We have confirmed again that
dependence on the choice of $\{ c_i \}_i$
is negligibly small.]
Not only 
$\beta(u; N)$ but also $\tilde{\beta}(u; \infty)$ 
depend on $N$.
However, 
the dependence of $\tilde{\beta}(u; \infty)$ becomes fairly weak
for $N \gtrsim 20$, as shown in the inset.
$\tilde{\beta}(u; \infty)$ for $N=24$ 
agrees well with
the exact result, 
whereas $\beta(u; N)$ differs significantly from them for this value of $N$.
We have thus confirmed that 
$\tilde{\beta}(u; \infty)$ is much closer 
to $\beta(u; \infty)$ than $\beta(u; N)$, for finite $N$.
Note however that $\beta(u;N)$ gives almost correct result for $\beta$
of a {\em finite} system, as seen from Eq.~(\ref{eq:betaN_better}).

We have obtained a series of TPQ states 
at the discrete points $u_0, u_1, u_2, \cdots, u_{\rm term}$.
The discrete points are dense enough because their intervals 
are $O(1/N)$, vanishing as $N \rightarrow \infty$.
The intervals also depend on the parameter $l$.
When smaller $l$ is taken, $k$ gets smaller to reach 
the same $u$ and temperature $T$, 
as seen from Eq.~(\ref{Nb=k/L}).
Hence, to obtain results at low $T$, 
$l\simeq e_{\rm max}$ is appropriate to reduce the amount of 
computation.
At high $T$, however, $u_{k}$ moves quickly as $k$ increases, 
for such a small $l$. 
Hence, to obtain results at many values of $u$ at high $T$, 
$l$ should be taken larger. 
When computing the data for Figs.~\ref{fig:hM} and \ref{fig:Cor},
we have taken $l \simeq e_{\rm max}$.
Since the values of $u$ which are specified in these figures 
are not necessarily found among $u_k$'s,
we have slightly tuned $l$ in such a way that 
$u_k$ can be found within $0.001J$ of the specified values.
In these figures, 
$m_z$ and $\phi(j)$ at such $u_k$'s are plotted.
When computing the data for Fig.~\ref{fig:ET}, 
we have performed computations with two values of $l$; 
$l=e_{\rm max}$ and $5J$.
Both results agree well with each other.
For better visualization, 
we have plotted the results with $l=5J$ (orange and purple) 
and those with $l=e_{\rm max}$ (red and blue) 
in the high- and low-$T$ regions, respectively.

{\em Advantages -- }
We now discuss advantages of our formulation 
when used as a method of numerical calculation.
At finite $T$, 
an exponentially large number of states are included in $\mathcal{E}_{E,N}$.
This makes computation of eigenstates in $\mathcal{E}_{E,N}$ pretty hard.
In contrast, 
our method takes full advantage of such a huge number of states,
as seen, e.g., in the derivation of Eq.~(\ref{Sd=e^xi}).
As a result, 
using just a single realization of TPQ state, 
one can calculate all quantities of statistical-mechanical 
interest at finite $T$, on the solid theoretical basis that 
is developed in this Letter.
Moreover, our method is applicable to 
systems of any spatial dimensions, and to 
frustrated or fermion systems as well.
Furthermore, 
our method costs much less computational resources than 
the numerical diagonalization.
For example, 
the number of non-vanishing elements of 
$\hat{H}$ of the Heisenberg model is
$O(N 2^N)$. 
Since $k=O(N)$, 
the computational time is $O(N^2 2^N)$ in our method, 
which is exponentially shorter than that of diagonalization.
In fact, 
it took only two hours to compute all data in 
Fig.~\ref{fig:ET} 
on a PC.
Computations can be made even faster by parallelizing 
the algorithm, which is quite easy and efficient because our method  
consists only of matrix multiplications.

Furthermore, 
our method is effective over a wide range of 
$T$ 
because the rhs of Eq.~(\ref{DeltaA}) is exponentially small 
as long as $s$ (and hence $T$) is finite of $O(1)$.
In fact, 
Figs.~\ref{fig:hM}-\ref{fig:ET} show that 
our results agree well with the rigorous results in a wide range of $T$, 
from $T \ll J$ to $T \gg J$.
%
%
In practical computations with finite $N$,
$T$ ($=1/\tilde{\beta}(u_\kappa^{\bullet};\infty)$) 
can be lowered as long as $r_k(e_{\rm min};N)/r_k(u_\kappa^*;N) \ll 1$.

We note that the quantum Monte Carlo method may be much faster.
However, it suffers from the sign problem 
in frustrated systems and fermion systems.
The density-matrix renormalization group method
has been extended to finite temperature,
and the state obtained in Ref.~\cite{White}
might be close to TPQ states. 
However, its effectiveness in two- or more-dimensional systems 
is not clear yet.
The states obtained with the microcanonical Lanczos method \cite{Long},
which tried to obtain not TPQ states but eigenstates,
might also be close to TPQ states.
However, the method 
costs more computational time in Ref.~\cite{Long}
than ours, and 
a method of computing $T$ or $s$ 
seems more difficult than ours.
We therefore expect that our method will 
make it possible 
to analyze systems which could not be analyzed with other 
methods.

{\em Concluding remarks -- }
We conclude this Letter by making several remarks.
First,
one can evaluate the magnetic susceptibility 
$(\partial m_z / \partial h_z)_u$ from
Fig.~\ref{fig:hM} or \ref{fig:Cor}.
One can also obtain 
$(\partial m_z / \partial h_z)_T$ with the help of 
Fig.~\ref{fig:ET}. 

Second, 
$| \psi_{k} \rangle$ remains to be a TPQ state 
after time evolution, 
since Eq.~(\ref{psi=Sdn}) shows that 
$e^{\hat{H}t/{i\hbar}} | \psi_{k} \rangle
\propto
\sum_{n} e^{- ie_n t/\hbar} c_{n} (l-e_{n})^k |n \rangle
$, 
which is just another realization of $| \psi_{k} \rangle$.

Third, 
our formulation is advantageous to analyses of phase transitions.
As an example, consider the case where the energy density 
$u(T;N)$ for $N \to \infty$ is discontinuous at 
the transition temperature $T_{\rm tr}$ of a 
first-order transition.
Then, the specific heat $c = {\partial u / \partial T}$
diverges at $T=T_{\rm tr}$.
If one used the canonical formalism, 
where $T$ is an independent variable, 
calculation of 
$u(T;\infty)$ would be hard around $T = T_{\rm tr}$.
In our formulation, by contrast, 
$u$ is taken as an independent variable, 
and $c$ is obtained as
$c 
=-\beta^2/({\partial \beta / \partial u})_{\scriptscriptstyle N}$
from $\beta(u;\infty)$. 
The function $\beta(u;\infty)$
is {\em continuous even at the transition point},
where it takes a constant value $1/T_{\rm tr}$
in a finite interval of $u$  
corresponding to the phase coexistent region \cite{AS:phaserule}.
Hence, $\beta(u;\infty)$ can be calculated more easily than $u(T;\infty)$. 
In fact, one can identify a first-order transition 
by simply observing that 
the rhs of Eq.~(\ref{eq:beta_infty}) takes a constant value,
apart from small deviation of $O(1/N^2)$,
%
%
for multiple values of $\tilde{u}^{\bullet}_\kappa$ and $\kappa$.
Regarding a continuous transition, 
it can be identified from a singularity in $c$, or
an order parameter $m$, and so on.
One can calculate $m$ by adding  
a symmetry-breaking field $f$ to the Hamiltonian,
and thereby computing $m$ at $f= \pm |f|$ for small $|f|$.
Or alternatively, without introducing $f$, one can perform the 
`pure-state decomposition' (i.e., decomposition into 
macroscopically definite states) by applying
the variance-covariance matrix method of Ref.~\cite{VCM} to a TPQ state. 

Fourth, 
Eq.~(\ref{psi=Sdn}) can be generalized as
$ 
|\psi \rangle \propto Q(\hat{h}) |\psi_{0} \rangle,
$ 
which defines other new TPQ states.
Here, $Q(u)$ is any differentiable real 
function such that $Q(u)^2 g(u;N)$ has a sharp peak,
whose width vanishes as $N \to \infty$,
and $Q(u)^2 g(u;N)$ outside the peak decays quickly. 
Using this $|\psi \rangle$,
one can calculate various quantities as we have done using $|\psi_k \rangle$.
For instance, the formula corresponding to 
Eq.~(\ref{Nb=k/L}) is given by
$
\beta(u^{\ast};N) 
+ 2Q'(u^{\ast}) / N Q(u^{\ast}) = 0
$.

Finally, 
although a TPQ state (such as $| \psi_k \rangle$) and 
the mixed state (such as $\hat{\rho}_k$) of the corresponding ensemble
are identical with respect to mechanical variables, 
{\em they are completely different with respect to entanglement}.
At $T \gg J$, for example, $\hat{\rho}_k$ has only small 
entanglement (because it is close to the completely mixed state 
$(1/D) \hat{1}$, 
which has no entanglement), whereas we can show 
that $| \psi_k \rangle$ has exponentially 
large entanglement (as previously shown for $T \to \infty$ in Ref.~\cite{SugitaShimizu}). 
It is thus seen that an equilibrium state can be represented 
either by a TPQ state with huge 
entanglement or by a mixed state with much less entanglement.
Their difference can be detected only by high-order polynomials of
local operators, which are 
{\em not} of statistical-mechanical interest \cite{SugitaJ, SugitaShimizu}.

\begin{acknowledgments}
We thank 
J. Sato, F. G\"{o}hmann, C. Trippe and 
K. Sakai for providing us with numerical
data of exact solutions.
We also thank H. Tasaki, A. Sugita, Y. Kato, Y. Oono, H. Katsura, 
K. Hukushima, S. Sasa and T. Yuge
for helpful discussions.
This work was supported by KAKENHI Nos.~22540407 and 23104707.
\end{acknowledgments}


\begin{thebibliography}{99}
\bibitem{vNS}
Entropy could be obtained
if one could obtain a TPQ state of a huge system which includes the target system. 
However, this is harder than calculating the partition function.
\bibitem{en:Boltzmann}
These are the conditions that the Boltzmann formula gives the correct 
thermodynamic entropy and that the system is stable.
Hence, they are necessary for all microscopic 
models to which statistical mechanics is applied.
\bibitem{norm}
When $\hat{A}$ is unbounded, 
$\| \hat{A} \|$ should be replaced with 
$
\max_{|\phi \rangle \in \mathcal{H}_N^\subset}
|\langle \phi | \hat{A}| \phi \rangle |
$,
where $\mathcal{H}_N^\subset$ denotes a Hilbert subspace
in which the values of macroscopic variables (such as $u$ and $m_z$) 
are limited to certain finite ranges.
\end{thebibliography}

\end{document}